\documentclass[aps,prl,preprint,groupedaddress]{revtex4-1}
\usepackage{bbm}
\usepackage{natbib}
\usepackage{graphicx}
\usepackage{appendix}
\usepackage{amsmath}	
\usepackage{amsfonts}	
\usepackage{amssymb}	
\usepackage{amsthm}	    
\begin{document}


\title{\textbf{Ansatz for the Jahn-Teller triplet instability}}
\author{ Arnout Ceulemans}
\affiliation{{Department of Chemistry, Katholieke Universiteit Leuven, \\ Celestijnenlaan 200F, B-3001 Leuven, Belgium, \\ arnout.ceulemans@kuleuven.be}}
\date{\today}
\begin{abstract}
A threefold degenerate electronic state is Jahn-Teller unstable with respect to symmetry lowering distortions, which transform as the five quadrupolar modes. The solution of the corresponding vibronic Hamiltonian is constructed using the analytical method introduced by Bargmann, as an alternative to existing group-theoretical methods based on coefficients of fractional parentage. It involves the construction of an Ansatz which incorporates SO(5) to SO(3) symmetry breaking. The resulting Jahn-Teller equations are derived, and solved in terms of radial polynomials and Gegenbauer functions.
\end{abstract}
\maketitle
\section{Introduction}
Both physics and chemistry abound with properties that have a Jahn-Teller instability at their core \cite{Bersuker}.
In turn the archetypal Jahn-Teller Hamiltonians offer valuable model systems for the study of fundamental Quantum Theory \cite{Englman}. They describe the coupling between a degenerate electronic system of fermions with an harmonic array of bosons. 

Different epochs in quantum mechanics have made use of different ways to present harmonic oscillators. In the Schrödinger representation the harmonic oscillator is described by the Hermite second-order differential equation. Dirac introduced the creation and annihilation operators which define an algebra that removed the need to solve differential equations. In a further development Bargmann went one step further and replaced the creation and annihilation operator by a complex variable, and its conjugate derivative \cite{Bargmann}. This Bargmann-Fock transformation reconnects the oscillator to the treasure chest of classical analysis, be it now at the level of first-order differential equations.

Application of this transformation greatly enriches our understanding of Jahn-Teller Hamiltonians. It was applied successfully to the single mode Rabi model, and the dual mode $E\times e$ Jahn-Teller model \cite{Braak, Reik, Ceulemans1,Ceulemans2}. It can be applied as well to the instabilities of the quartet spinor $\Gamma_8$.
These systems share an important symmetry characteristic which considerably facilitates the treatment.
In both Jahn-Teller cases there is a common maximal orthogonal Lie group which covers both the fermion and boson space. In the case of $E\times e$ this is the cyclic symmetry group SO(2). The $e$-bosons are the fundamental vector representation of SO(2), with angular momentum $m_{\Lambda} = \pm 1$, while the $E$-fermion system transforms as the fundamental spinor representation $m_{\Lambda} = \pm 1/2$. In the $\Gamma_8\times t$ system the symmetry group is spherical symmetry SO(3), where the $t$-bosons again form the fundamental $L=1$ representation, and the $\Gamma_8$ can be resolved into two pseudo-spins  $S=1/2$ \cite{Judd1}. In the $\Gamma_8 \times (e+t_2)$ problem under equal coupling conditions the symmetry group is the five-dimensional hypersphere, SO(5). The five boson modes describe the $(1,0)$ fundamental vector representation of this symmetry group. SO(5) has the peculiarity that it is isomorphic to the symplectic group Sp(4). This group has a four-dimensional spinor at its core which precisely describes the fermion quartet.

Not so for the $T\times(e+t_2)$ problem. Here, as in the case of the $\Gamma_8 \times(e+t_2)$, the nuclear coordinate space under equal coupling furnishes the fundamental vector for the SO(5) hypersphere, but this symmetry group lacks a threefold degenerate representation to accommodate the fermion triplet. The maximal symmetry of the fermion triplet is the $L=1$ vector of SO(3). The symmetry of the Jahn-Teller Hamiltonian is thus limited to SO(3). This means that the interaction represents a symmetry breaking: $SO(5) \downarrow SO(3)$. 

The coefficients of fractional parentage that describe this subduction process have been used to construct a huge interaction matrix, the secular equation of which yields the dynamical levels of the $T\times(e+t_2)$ Hamiltonian. The vibronic levels with $L=1$ symmetry have been derived by O'Brien \cite{Brien} and have been discussed in great detail by Judd \cite{Judd2}. Interestingly exactly the same problem appears in nuclear physics, where the lowest dipole transition is obtained through the coupling of the dipole transition moment to the quadrupole surface vibrations of the nucleus \cite{Tourneux}. The group-theoretical construction of interaction matrices requires extensive sets of coupling coefficients and relies on delicate phase conventions. Nowadays these treatments are superseded by large scale diagonalization in the full harmonic oscillator basis \cite{Naoya}.

The present contribution aims to derive the Ansatz for the triplet Jahn-Teller problem under equal coupling conditions. This Ansatz gives rise to a set of coupled first-order Jahn-Teller differential equations, which express this fermion-boson coupling scheme according to the Bargmann representation. This is uncharted territory which may offer a new perspective on this intriguing case of symmetry breaking.
\section{The Hamiltonian}
Let $\mathbf{f}^\dagger$ create the three states of the fermion, which transform as the $T$ representation in the cubic and icosahedral groups, and correspond to $L=1$ in spherical symmetry. The fermion components are labeled by the $m_l$ values, as $f^\dagger_{+1}, f^\dagger_{0}, f^\dagger_{-1}$.
The Jahn-Teller modes  transform as the non-totally symmetric part of the symmetrized square of the fermionic symmetry representation. In spherical symmetry, this part corresponds to the 
$L=2$ quadrupole representation. The fermionic quadrupole operator is given by:
\begin{eqnarray}
\left( \mathbf{f}^{\dagger} \mathbf{f} \right)^2_q &=&\sum _{m_1m_2} (-1)^{1-m_2} {\langle} 1m_1 1m_2|2q{\rangle} f^\dagger_{m_1} f_{-m_2} \nonumber \\
&=& \sum _{m_1m_2} (-1)^{1-m_2} (-1)^q \sqrt{5}
\left( \begin{matrix} 1&1&2 \\m_1& m_2 & -q \end{matrix} \right) f^\dagger_{m_1} f_{-m_2}
\end{eqnarray}
where the coupling is performed by the 3J symbol.
This problem is also designated as the $P\times d$ problem. 
In octahedral groups the $L=2$ tensor spans two irreducible representations (irreps), viz. $e_g + t_{2g}$, while in the icosahedral case it subduces a single five-fold degenerate irrep, commonly denoted as the $h_g$-representation.
The present treatment is based on the so-called spherical or degenerate coupling case, where the five boson modes are degenerate and so are the vibronic coupling constants.
The coordinates in the usual octahedral form are the $e$-modes, $Q_{\theta}, Q_{\epsilon}$ and the $t_2$-modes $Q_{\xi}, Q_{\eta},Q_{\zeta}$. In this treatment we will use the complex basis set, defined as:
\begin{eqnarray}
Q_0 &=& Q_{\theta} \nonumber \\
Q_{+1} &=& -\frac{1}{\sqrt 2} Q_{\eta} - \frac{i}{\sqrt 2} Q_{\xi} \nonumber \\
Q_{-1} &=& \frac{1}{\sqrt 2} Q_{\eta} - \frac{i}{\sqrt 2} Q_{\xi} \nonumber \\
Q_{+2} &=& \frac{1}{\sqrt 2} Q_{\epsilon} + \frac{i}{\sqrt 2} Q_{\zeta} \nonumber \\
Q_{-2} &=& \frac{1}{\sqrt 2} Q_{\epsilon} - \frac{i}{\sqrt 2} Q_{\zeta}
\label{complexQ}
\end{eqnarray}

The Hamiltonian consists of a harmonic part, $\mathcal{H}_0$, and a linear coupling term, $\mathcal{H}'$, which corresponds to the scalar product of the fermion tensor and the coordinate tensor. 
\begin{equation}
\mathcal{H}'= k \sum_q (-1)^q ( \mathbf{f}^{\dagger} \mathbf{f} )^2_q Q_{-q}
\end{equation}
Here k is the coupling parameter. This product may be expressed in matrix form, acting in the space of the fermions, ordered as $|+1\rangle,|0\rangle,|-1\rangle$.
\begin{equation}
\mathcal{H}' =
k\left( 
\begin{matrix}
\frac{1}{\sqrt{6}} Q_0 & \frac{1}{\sqrt{2}} Q_{-1} & Q_{-2} \\
-\frac{1}{\sqrt{2}}Q_{+1} & -\frac{2}{\sqrt{6}}Q_0 &  -\frac{1}{\sqrt{2}} Q_{-1}\\
 Q_{+2} & \frac{1}{\sqrt{2}}Q_{+1} & \frac{1}{\sqrt{6}} Q_0
\end{matrix}
\right)
\end{equation}
As this is a scalar product of spherical tensors, the coupling Hamiltonian will be SO(3) invariant.
It is  instructive to deal first with the so-called static Jahn-Teller problem, as opposed to the dynamic problem which includes the nuclear vibrations \cite{Chancey}. 
The secular equation of $\mathcal{H}'$ reduces to:
\begin{equation}
E^3 - \frac{E}{2} Q^2 +\frac{1}{3\sqrt{6}} I_3^3 =0
\end{equation}
where:
\begin{eqnarray}
Q^2 &=& Q^2_0 - 2Q_{+1}Q_{-1} + 2 Q_{+2}Q_{-2} \nonumber \\
I_3 &=& Q_0 ( Q_0^2-6 Q_{+2}Q_{-2}- 3  Q_{+1}Q_{-1}) +\frac{3\sqrt{3}}{\sqrt{2}}\left(  Q_{+2} Q^2_{-1}+Q_{-2}Q^2_{+1}\right)
\end{eqnarray}
Since $\mathcal{H}'$ is traceless, the quadratic term in this equation is absent. This case is called a depressed cubic. The secular equation contains two SO(3) invariants: $Q$ is the radius of the distortion, while $I_3$ is a third-order invariant, proportional to the determinant of the JT Hamiltonian.
The roots of the eigenvalue equation can be expressed using the angle representation.
Rewrite $I_3$ as:
\begin{equation}
I_3= Q^3 \cos{3\gamma}
\label{I3}
\end{equation}
The equation can then easily be solved by trigonometric expressions for the three roots:
\begin{equation}
E_k = -k\frac{2Q}{\sqrt{6}}\cos\left({\gamma -\frac{2n\pi}{3}}\right) \; n=0,1,2
\end{equation}
These roots are linear in $Q$ and stand for the symmetry breaking terms, which cause the Jahn-Teller instability.The potential energy term in $\mathcal{H}_0$ is proportional to $Q^2$ and acts as the symmetry restoring potential, which limits the extent of the distortion.

Now to proceed to the dynamic problem, we introduce the quadrupole bosons. 
The vibrational modes are created by the tensor operator $\mathbf{b}^\dagger$, with components $b^\dagger_{+2}, b^\dagger_{+1}, b^\dagger_{0}, b^\dagger_{-1}, b^\dagger_{-2}$.
It must be realized that the associated annihilation operators have complex conjugate properties. 
While the creation of a boson with $m_l=+2$ component adds $+2$ to the total $M_L$ value of a given state, its annihilation will have the effect of lowering $M_L$ by two units. The $b_{+2}$ operator thus is characterized by $m_l=-2$. Taking also into account the phase conventions as specified in Eq. \eqref{complexQ}, the proper annihilation tensor, denoted by a tilde, is given by:
\begin{eqnarray}
\tilde{b}_0 &=& b_0 \nonumber \\
\tilde{b}_{+1} &=& - b_{-1} \nonumber \\
\tilde{b}_{-1} &=& - b_{+1} \nonumber \\
\tilde{b}_{+2} &=&  b_{-2} \nonumber \\
\tilde{b}_{-2} &=&  b_{+2}
\label{tilde} 
\end{eqnarray}
The coupling Hamiltonian for the triplet case can then be rewritten as:
\begin{equation}
\mathcal{H}' =  \kappa \left( \mathbf{f}^{\dagger} \mathbf{f} \right)^2  \odot \left(\mathbf{b}^\dagger + \mathbf{\tilde{b}}\right)^2
\end{equation}
with $\kappa = k/\sqrt{2}$.
%
%
%
\section{The five-dimensional harmonic oscillator}
\subsection{Quadrupolar distortions of a sphere}
The surface deformations of a sphere are characterized by the symmetries of the spherical harmonics. The $L=0$ mode is of course isotropic and unique, and corresponds to a radial breathing mode. The next $L=1$ dipolar symmetry characterizes a spurious translation mode. The $L=2$ quadrupolar symmetry thus corresponds to the first non-isotropic deformation of a sphere \cite{Eisenberg, Ceulemans3}. Since this is the smallest possible non-zero $L$ value, the quadrupolar modes can only introduce a minimal symmetry breaking: they distort a sphere into an ellipsoid. An ellipsoid is a surface characterized by three orthogonal axes of different length. The sum of these lengths must be constant in time, in order to avoid any admixture of the radial breathing mode. Hence proper ellipsoidal distortions have only two degrees of freedom. These correspond to the tetragonal $Q_\theta$ mode and the orthorhombic $Q_\epsilon$ mode. The tetragonal mode leads to a prolate or oblate ellipse, which still has cylindrical symmetry along the z-axis. 
The radius of this ellipse is thus described as:
\begin{equation}
d(\theta,\phi) = R\left( 1 + c (3 \cos^2\theta - 1)\right)
\end{equation}
where R is the radius of the sphere, and c is scaling constant which oscillates in time with the vibration. According to this expression the distortions along the three Cartesian directions are as follows:
\begin{eqnarray}
d(0,\phi)-R &= 2cR \nonumber \\
d({\pi} /2, 0)-R &= -cR \nonumber \\
d({\pi} /2,{\pi} /2)-R &= -cR
\end{eqnarray}
The orthorhombic mode will further break this axial symmetry, by repartitioning the distortion between the $x$- and $y$- directions. A general ellipsoidal distortion with principal axes along the Cartesian directions is thus described by a vector in the space formed by these coordinates. Turning to polar coordinates, the parametric description of this distortion reads:
\begin{equation}
\left(
\begin{matrix}
Q_\theta \\ Q_\epsilon\\ Q_\xi \\ Q_\eta \\Q_\zeta
\end{matrix} \right)
= Q
\left(
\begin{matrix}
 \cos{\gamma} \\
\sin{\gamma} \\ 0 \\ 0 \\ 0
\end{matrix}
\right)
\end{equation}
or, in the case of the complex harmonics:
 \begin{equation}
\left(
\begin{matrix}
Q_{+2} \\Q_{+1}\\ Q_0 \\Q_{-1}\\ Q_{-2}
\end{matrix} \right)
= Q
\left(
\begin{matrix}
\frac{1}{\sqrt{2}} \sin{\gamma} \\ 0 \\ \cos{\gamma} \\ 0 \\
 \frac{1}{\sqrt{2}} \sin{\gamma}
\end{matrix}
\right)
\label{15}
\end{equation}
Note that the angle in these equations is indeed the angle $\gamma$, which we used in Eq. \eqref{I3} to express the determinant of $\mathcal{H}'$, as can easily be verified by substitution.

The quadrupole distortion mode is bimodal, since it is characterized by two radial degrees of freedom. The ellipsoid which is obtained in this way is still aligned with the Cartesian reference frame. Spherical symmetry of course requires that the ellipsoid is free to rotate in 3D space. This is where the three remaining quadrupolar modes come in. The general orientation of the ellipsoidal distortion can be performed by the Euler rotation matrix in the full space of the five $L=2$ modes. In the usual way the coordinates are presented as a column vector, as in Eq. \eqref{15}. Now let $Q$ be the radius of the distortion space, then the general angular form of the coordinates is given by:
\begin{equation}
Q_M = Q \left[ \cos \gamma D_{M0} + \frac{\sin \gamma}{\sqrt{2}} \left(D_{M2} + D_{M-2} \right)\right]
\label{16}
\end{equation}
Here $D_{ij}$ are elements of the angular overlap matrix.
Let us summarize what we have got: the electronic triplet couples to five modes, which transform as the $L=2$ spherical harmonics. This explains the overall rotational symmetry of this Jahn-Teller surface, as described by the SO(3) group. The coordinate space can be expressed by two radial distortion modes, which create the ellipsoid, and three angles which orient the ellipsoid in 3D space.

As far as the Jahn-Teller force is concerned, in the lowest energy through it is directed towards the prolate or oblate ellipse, corresponding to the $\gamma = 0$ direction. Distortions along the second mode correspond to higher energy blades of the surface. This means that there is no free rotation over the angle $\gamma$, as would be the case for a degenerate bimodal regime.
Therefore the Jahn-Teller surface is described by a bimodal distortion, which creates the ellispoid, and three angles which orient it in 3D space. From these three angles the surface takes its SO(3) symmetry, but that is it.

In contrast, for the vibrating sphere the quadrupole modes are properly be described by only one radial coordinate, and four free angles. This coordinate space thus corresponds to a 4-sphere in 5D. The symmetry of this space is thus SO(5). It is larger than the Jahn-Teller symmetry, while on the other hand it is only a subgroup of the full symmetry of the quadrupole oscillator, which is SU(5). From the moment that the Jahn-Teller coupling sets in, symmetry is entirely broken from SU(5) to SO(3). The five dimensional rotation group SO(5) plays an important role as an intermediate subgroup in this chain:
\begin{equation}
SU(5)  \rightarrow SO(5) \rightarrow SO(3)
\end{equation}
The SU(5) group corresponds to the oscillator Hamiltonian at zero coupling. Restriction of SU(5) to its subgroup SO(5) is achieved by considering the real coordinates of the distortion space. Finally the SO(3) symmetry is what remains after the Jahn-Teller coupling is activated.
%
%
\subsection{Boson representations in SO(5)}
The spectral analysis of the subduction chain from SU(5) to SO(3) has been described by Judd \cite{Judd2}. Extensive treatments are available in the nuclear physics literature \cite{Chacon, Rowe1,Rowe2}. The $n$-fold degenerate levels of the 5D harmonic oscillator form irreps of SU(5), denoted simply by the counting number $[n]$. The relevant irreps of SO(5) are all of $(\nu,0)$ type, where $\nu$ is Racah's seniority number.
In SO(3) irreps are labeled by the $L$ quantum number. 
The dimensions of these representations for the three groups involved are as follows, with ${l}=2$:
\begin{eqnarray}
\dim[n] &=& (2l+1) (2l+2) ...(2l+n)/n! \nonumber \\
\dim(\nu,0) &=& (\nu + 1)(\nu+2)(2\nu+3)/6 \nonumber \\
\dim(L) &=& 2L+1
\end{eqnarray}
The branching rules from SU(5) to SO(5) are as follows:
\begin{eqnarray}
[2n] &\rightarrow& (2n,0) + (2n-2,0) + ... + (0,0) \nonumber \\
{[2n+1]} &\rightarrow& (2n-1,0) + (2n-3,0) + ... + (1,0)
\end{eqnarray}
 As for the branching rules from SO(5) to SO(3), we will address here only the lowest $L$ states,
 with $L\leq 3$.
Quite remarkably, there are no $P$ states. Furthermore, the branching shows a perfect repetition with a period of three. Let $\mu =\nu\mod3$:
\begin{eqnarray}
\mu=0 &:& (\nu,0) \rightarrow S \nonumber \\
& & (\nu+3,0) \rightarrow F \nonumber \\
\mu = 1 &:& (\nu,0) \rightarrow D \nonumber \\
\mu=2 &:& (\nu,0) \rightarrow D
\end{eqnarray}
Figure 1 offers a 2D plot of the spectrum of $S$, $D$, and $F$ states, with oscillator energy, denoted by the SU(5) label $[n]$, on the vertical axis, and the SO(5) label $(\nu,0)$ on the horizontal axis. The pattern is periodic in $3\nu$. The ground level $[0]$ is the vacuum state. It has a zero-point energy of $5/2 \hbar\omega$. This is usually taken as the zero of energy, and $\hbar \omega$ is taken as the unit of energy.
\begin{figure*}
        \includegraphics*[width=\columnwidth]{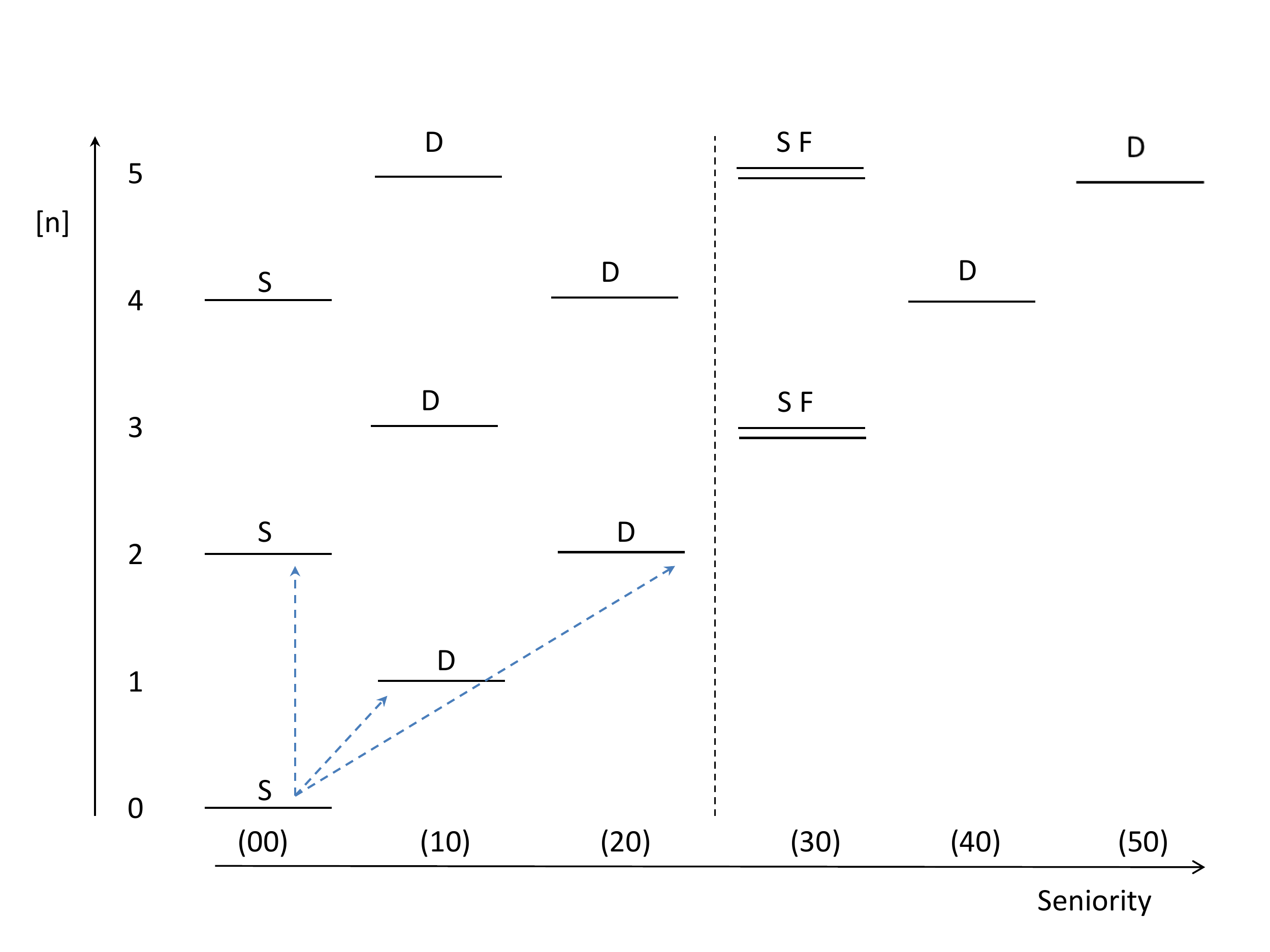}
        \caption{Harmonic oscillator states with $L\leq 3$ as a function of $[n]$ ($n\leq 5$) and seniority. }
\end{figure*}
The first boson excitation reaches the $(1,0)$ level which is five-fold degenerate and constitutes the basic vector of the hypersphere. In spherical symmetry it corresponds to the D-representation. The creation operators have the same symmetry, and thus the same angular dependence as the Q-coordinates. We can thus rewrite Eq. \eqref{16} as:
\begin{equation}
b^{\dagger}_M = \beta^{\dagger} \left[ \cos \gamma D_{M0} + \frac{\sin \gamma}{\sqrt{2}} \left(D_{M2} + D_{M-2} \right)\right]
\label{b}
\end{equation}
Here $\beta^\dagger$ is associated with the creation of a radial excitation in SO(5).

Now let us consider the 15 states that results from the excitation of two bosons.
These states can conveniently be represented in a Cartan-Weyl diagram \cite{Judd1}. An introduction to the use of these diagrams can be found in ref. \cite{Ceulemans4}. In the case of the SO(5) Lie algebra, individual states can be made to be eigenfunctions of two commuting operators. 
These operators can be expressed as follows:
\begin{eqnarray}
\mathcal{L}_1&=& b^{\dagger}_{+2}b_{+2}-b^{\dagger}_{-2}b_{-2} \nonumber \\ 
\mathcal{L}_2&=& b^{\dagger}_{+1}b_{+1}-b^{\dagger}_{-1}b_{-1}
\end{eqnarray}
They are at the basis of a two-dimensional Cartan-Weyl diagram. 
Eigenvalues of the states with respect to these two operators address a point in the diagram, as shown in figure 2. 
As an example for the $(b^\dagger_{+2})^2$ boson, one has the following commutation relations: 
\begin{eqnarray}
 \left[\mathcal{L}_1,(b^\dagger_{+2})^2\right]&=& 2 (b^\dagger_{+2})^2\nonumber \\
\left[\mathcal{L}_2,(b^\dagger_{+2})^2\right]&=& 0
\end{eqnarray}
In the $\mathcal{L}_1, \mathcal{L}_2$ diagram it thus has coordinates $(2,0)$.
In this way 15 two-boson states can be placed in the diagram of Figure 2.
The SO(3) angular momentum operator which measures the total $M_l$ value, recognizes the bosons by their $m_l$ value as follows:
\begin{equation}
\mathcal{L}_z = 2 b^{\dagger}_{+2}b_{+2}+b^{\dagger}_{+1}b_{+1}- b^{\dagger}_{-1}b_{-1} -2 b^{\dagger}_{-2}b_{-2}=2\mathcal{L}_1 + \mathcal{L}_2
\end{equation}
\begin{figure*}
        \includegraphics*[width=\columnwidth]{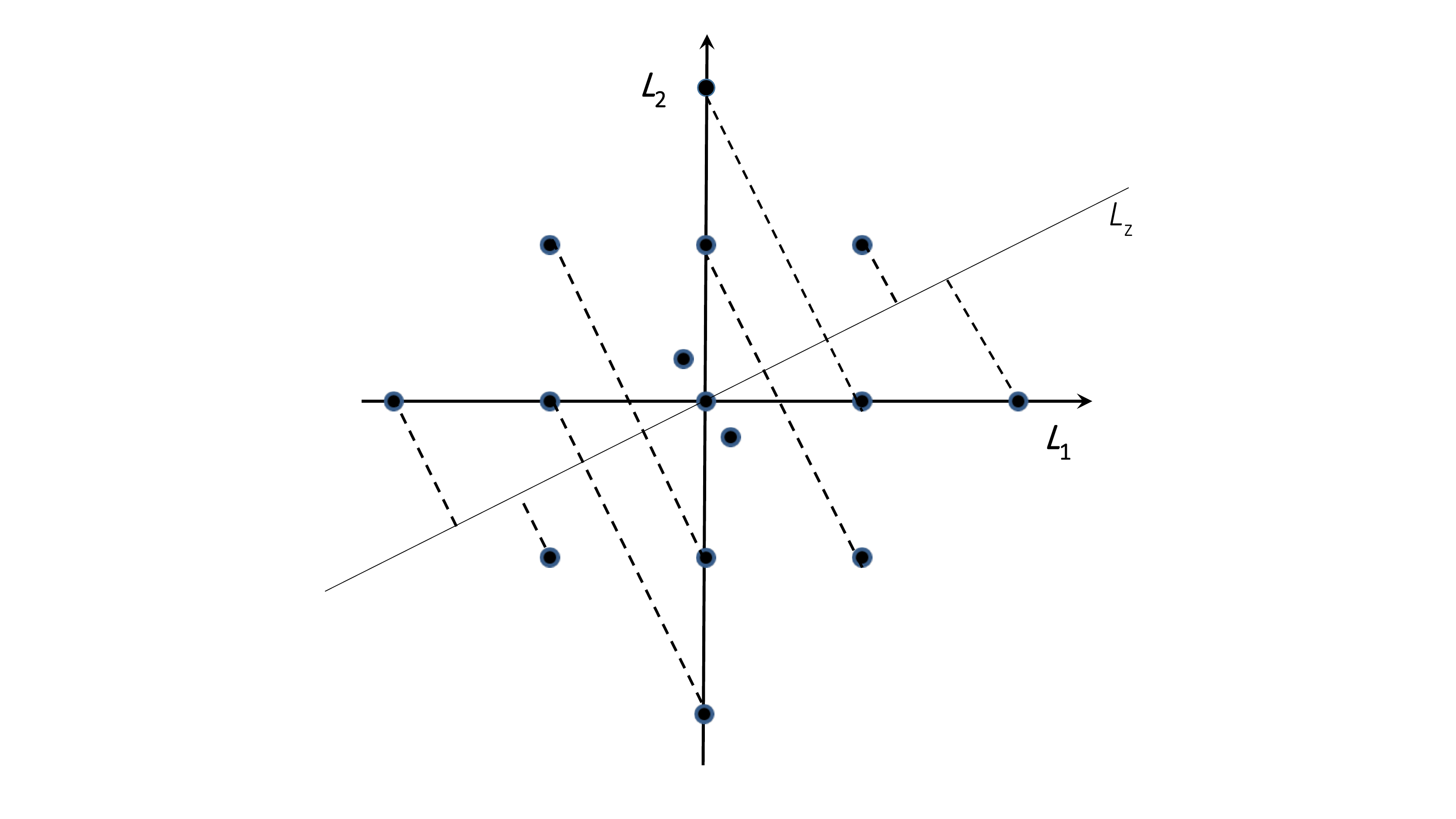}
        \caption{Display of the 15 two-boson states in the Cartan-Weyl diagram of SO(5).}
\end{figure*}
In the diagram $\mathcal{L}_z$ values can thus be directly read off by projecting the components on a straight line, making an angle of $\arctan(1/2) \approx 26.56^\circ$ with the $\mathcal{L}_1$ axis. Scaling must be adapted accordingly, i.e. equidistant points on this line are one unit of $m_l$ values apart. The diagram then reveals that the projections of the fifteen boson modes can be partitioned into three sets of equidistant points, with cardinalities $9, 5, 1$. Our space thus subduces $G+D+S$ multiplets in SO(3). Further, the top coordinate of the diagram is used to provide a label for the SO(5) state, hence the diagram contains the $(2,0)$ representation of SO(5). This representation is 14-dimensional, hence it will subduce $G+D$. The remaining singleton is the isotropic S state, corresponding to $(0,0)$ in SO(5).
The branching scheme is thus as follows:
\begin{equation}
[2] \rightarrow 
\left\{
\begin{matrix}
 (2,0) &\rightarrow& D+G  \\
(0,0) &\rightarrow &S
\end{matrix}\right .
\end{equation}
In order to obtain explicit expresssions for the SO(3) multiplets, the tools of standard angular momentum theory can be put at work.
 The isotropic S state of course corresponds to the creation of two radial quanta from the S type vacuum state. It is thus given by:
\begin{equation}
(\beta^{\dagger})^2 = 2 b^\dagger_{+2}b^\dagger_{-2}
- 2 b^\dagger_{+1}b^{\dagger}_{-1} 
 + (b^\dagger_0)^2
\end{equation}
Furthermore since the $M_L=\pm4$ and $M_L=\pm3$ substates occur only once, they are identified at once as the $|G\pm4\rangle$ and $|G\pm3\rangle $ substates.
A straightforward way to derive the $D$ state is by using the appropriate coupling coefficients:
\begin{eqnarray}
 (\mathbf{b}^\dagger \mathbf{b}^\dagger )^2_{+2}&=& \sum_{m_1m_2} (-1)^{+2} \sqrt{5}
\left( \begin{matrix}
2 & 2 & 2 \\
m_1 & m_2 & -2
\end{matrix} \right)
{b}^\dagger_{m_1} {b}^\dagger_{m_2}\nonumber \\
&=& \frac{1}{\sqrt{7}} \left(2\sqrt{2} b^\dagger_{+2}b^\dagger_0 -\sqrt{3} (b^\dagger_{+1})^2 \right)
\end{eqnarray} 
This formula produces the correct substate, except for normalization.
This is a delicate aspect related to the difference between boson and fermion couplings.
When applying the 3J coupling formula, the first term is produced twice, once as $b^\dagger_{+2}b^\dagger_0$ and once as $b^\dagger_0 b^\dagger_{+2}$. For bosons these two terms are identical. 
This increases the weight of this term by a factor of 2. The second term in  $(b^\dagger_{+1})^2$ appears only once in the coupling, but its weight is
also increased by a factor of 2, due to the boson normalization rule. As a result the $\sqrt{7}$ in the equation should be replaced by a $\sqrt{14}$, to obtain a properly normalized boson combination.

From here we will stick to the following expressions for the two-boson D states to be in line with the treatment by Gheorghe et al., which will serve as a basis to define our Ansatz \cite{Gheorghe}.
\begin{eqnarray}
\left(\mathbf{b}^\dagger \mathbf{b}^\dagger \right)_{\pm 2}^2 &=& \frac{\sqrt{3}}{\sqrt{2}} b^\dagger_{\pm 1} - 2 b^\dagger_{\pm 2}b^\dagger_{0} \nonumber \\
\left(\mathbf{b}^\dagger  \mathbf{b}^\dagger\right)_{\pm 1}^2 &=& b^\dagger_0 b^\dagger_{\pm 1} - \sqrt{6} b^\dagger_{\pm 2}b^\dagger_{\mp 1} \nonumber \\
\left(\mathbf{b}^\dagger  \mathbf{b}^\dagger\right)_{0}^2 &=& (b^\dagger_0)^2  -2 b^\dagger_{+2} b^\dagger_{-2} - b^\dagger_{+1}b^\dagger_{-1}
\label{bb}
\end{eqnarray}
Upon inserting the bimodal distortions the column vector of these two-boson modes becomes:
\begin{equation}
\left(
\begin{matrix}
\left(\mathbf{b}^\dagger  \mathbf{b}^\dagger\right)_{+ 2} \\ \left(\mathbf{b}^\dagger  \mathbf{b}^\dagger\right)_{+1}\\ \left(\mathbf{b}^\dagger  \mathbf{b}^\dagger\right)_{0\;\;\;} \\ \left(\mathbf{b}^\dagger  \mathbf{b}^\dagger\right)_{- 1}\\ \left(\mathbf{b}^\dagger  \mathbf{b}^\dagger\right)_{-2}
\end{matrix} \right)
= 
(\beta^{\dagger})^2 \left(
\begin{matrix}
-\frac{1}{\sqrt{2}} \sin{ 2\gamma} \\ 0 \\ \cos{2\gamma} \\ 0 \\
- \frac{1}{\sqrt{2}} \sin{2\gamma}
\end{matrix}
\right)
\end{equation}
Under spherical rotations this two-boson state transforms in exactly the same way as a quadrupole, since it also has D symmetry. Hence the general vector is obtained by applying the same angular overlap matrix:
\begin{equation}
\left(\mathbf{b}^\dagger \mathbf{b}^\dagger\right)_{M}^2= (\beta^\dagger)^2 \left[ \cos 2\gamma D_{M0} - \frac{\sin 2\gamma}{\sqrt{2}} \left(D_{M2} + D_{M-2} \right) \right]
\end{equation}

The isotropic three-boson S state in the $(3,0)$ representation also deserves special attention, since it is together with the radial boson a basic spherical invariant of our Jahn-Teller problem. It can be derived by coupling the singly and doubly excited D states. The standard notation which we will adopt for this state is $|S_{p,n}\rangle$, with $p=1$ and $n=0$. The subscripts in this notation refer to the position of this level in the diagram of Figure 1: the first subscript $p$ refers to the $(3p,0)$ level on
the horizontal axis of the diagram, while the second number $n$ is the number of radial quanta on the vertical axis, starting from the lowest S level at position $(3p,0)$. 
One has:
\begin{equation}
| S_{1,0}\rangle = b^\dagger_0 (-6b^\dagger_{+2}b^\dagger_{-2}- 3 b^\dagger_{+1}b^\dagger_{-1}+(b^\dagger_0)^2) + \frac{3\sqrt{3}}{\sqrt{2}}
\left(b^\dagger_{+2} (b^\dagger_{-1})^2 +b^\dagger_{-2} (b^\dagger_{+1})^2\right) 
\end{equation}
Upon substituting the polar forms of the operators in this expression one obtains:
\begin{equation}
|S_{1,0}\rangle = (\beta^{\dagger})^3 \cos {3 \gamma} 
\label{S3}
\end{equation} 
This invariant is clearly the operator expression of the $I_3$ invariant in the static problem.
%
%
%
%
\subsection{Bosonic eigenstates with $L \leq 3$}
Having explored the eigenspace of the harmonic quadrupole operator, the actual challenge now is to construct explicit harmonic functions for all the states shown in Figure 1. Several studies have been devoted to this, in view of the importance of the quadrupole oscillations as the spectral origin of the vibrating nucleus. The seminal paper on superspherical harmonics was written by Chac\'{o}n and Moshinsky. \cite{Chacon}. They provided generating functions for all SO(5) harmonics. These generating functions yield a complete set, but unfortunately the results are not orthogonal. To obtain the canonical forms one then has to recur to orthogonalization \cite{Rowe2}. This can of course be performed by standard computational techniques, but - in the absence of closed form formula - the non-orthogonal basis remains a major obstacle to the construction of the Ansatz.
Fortunately, for the present purposes, alternative approaches can be found in earlier works.
This line of inquiry was initiated by Bes, who derived some closed form formulas for states with angular momenta up to 6 \cite{Bes}. We will illustrate this approach for the simplest case of zero angular momentum in S states.
For S functions the angular momentum in the five-dimensional space is limited to rotations involving the $\gamma$ angle. The corresponding surface Laplacian operator \cite{Chancey} is given by:
\begin{equation}
\mathcal{O} = -\frac{1}{\sin{3 \gamma}} \frac{\partial }{\partial \gamma} \sin{3 \gamma} \frac{\partial }{\partial \gamma}
\end{equation}
The equation then reads:
\begin{equation}
\mathcal{O} |\Psi\rangle =  \Lambda |\Psi\rangle
\text{,           with:} \;\Lambda= \nu (\nu + 3)
\end{equation}
where $\nu$ is the seniority number which refers to the SO(5) representation $(\nu,0)$. As we have already determined, for S states this seniority label must be zero, or a multiple of 3.
Upon substituting $\nu=3p$,  $\Lambda$ becomes:
\begin{equation}
\Lambda = 9 p(p + 1)
\end{equation}

We now replace the angle $\gamma$ by the variable $x=\cos{3\gamma}$, which refers to the spherical invariant $I_3$.
One has:
\begin{equation}
\frac{\partial }{\partial \gamma} = -3 \sin{3\gamma} \frac{\partial }{\partial x} 
\end{equation}
The Laplace equation then becomes:
\begin{equation}
 (1-x^2) \frac{\partial^2 w}{\partial x^2}-2x\frac{\partial w}{\partial x} +p (p+1) w=0
\end{equation}
This is the Legendre equation in $x$, and the corresponding eigenvectors are the well known Legendre polynomials, $P_p(x) $.
These polynomials thus describe the $\gamma$-dependence of the S states. The radial part in turn is given by appropriate powers of the radial creation operator, addressing the ladder of S-states within a given seniority, $\nu=3p$. 
As illustrated in the oscillator diagram, the $|S_{p,n}\rangle$ state thus contains $3p + 2n$ radial quanta. 
\begin{equation}
|S_{p,n}\rangle =  (\beta^\dagger)^{3p+2n} P_p(x)
\end{equation}
The vacuum state, with $p=n=0$ corresponds to a constant. The angular part for the three-boson $|S_{1,0}\rangle$ state, as given in Eq. \eqref{S3}, is the Legendre polynomial of rank 1, which simply corresponds to $x=\cos{3\gamma}$.
Bes also derived recursion formulas for states with L up to 6, but failed to recognize that the results corresponded to simple linear combinations of Gegenbauer polynomials. These polynomials are in fact hyperspherical extensions of the Legendre polynomials, and as such are indeed suitable to describe harmonics on a sphere in 5D.
This connection was realized later in a most valuable paper by Gheorghe et al. who provided general closed form expressions \cite{Gheorghe}. These involve linear combinations of Gegenbauer polynomials, and form an ideal basis for the construction of an Ansatz. 
Gegenbauer polynomials are denoted as $C_p^{\lambda}(x)$, with $p$ referring to the rank and $\lambda$ is a dimensional parameter. These polynomials obey the following orthogonality relationship:
\begin{equation}
\int^{+1}_{-1} C_p^\lambda(x) C_q^\lambda(x) \left( 1-x^2\right) ^{\lambda-1/2} dx = 
\delta_{p,q} \frac{\pi \, 2^{1-2\lambda}\Gamma(p+2\lambda)}{p! (\nu+p) \left[ \Gamma(n)\right] ^2}
\end{equation}
In the case of a sphere in three dimensions, the parameter $\lambda$ is equal to $1/2$, and we obtain the well known Legendre polynomials:
\begin{equation}
P_{p}(x) = C^{1/2}_{p}(x)
\end{equation}
As indicated in Figure 1, for the $D$ states we have two spectral series: the series based on levels with seniority $\nu \mod{3} = 1$, and those with $\nu \mod{3} =2$. Gheorghe et al. describe the angular parts of these two series, apart from normalization, as:
\begin{equation}
\begin{split}
|\chi_{3p+1}^D M \rangle
= & C_p^{3/2} \left[  \cos \gamma D_{M0} + \frac{\sin \gamma}{\sqrt{2}} \left(D_{M2} + D_{M-2} \right) \right] \\
&-
 C_{p-1}^{3/2}\left[  \cos 2\gamma D_{M0} - \frac{\sin 2\gamma}{\sqrt{2}} \left(D_{M2} + D_{M-2} \right) \right] \\
|\chi_{3p+2}^D M\rangle =&- C_{p-1}^{3/2} \left[ 
  \cos \gamma D_{M0} + \frac{\sin \gamma}{\sqrt{2}} \left(D_{M2} + D_{M-2} \right) \right] \\
 & + C_p^{3/2} \left[  \cos 2\gamma D_{M0} - \frac{\sin 2\gamma}{\sqrt{2}} \left(D_{M2} + D_{M-2} \right) \right]
\end{split}
\end{equation}
The full states can now be conveniently represented by substituting the angular parts by the single and double boson excitations:
\begin{eqnarray}
|D^{1}_{p,n} M \rangle &=& (\beta^\dagger)^{3p+2n}  C_p^{3/2} (x) b^\dagger_M
-(\beta^\dagger)^{3p+2n-1}  C_{p-1}^{3/2}(x) (\mathbf{b}^\dagger\mathbf{b}^\dagger)_M \nonumber \\
|D^{2}_{p,n} M \rangle &=& -(\beta^\dagger)^{3p+2n+1}  C_{p-1}^{3/2}(x) b^\dagger_M
 + (\beta^\dagger)^{3p+2n} C_{p}^{3/2}(x) (\mathbf{b}^\dagger \mathbf{b}^\dagger)_M 
\end{eqnarray}
The lowest states are reached for $p=0$ and $n=0$. Since the rank of a Gegenbauer polynomial cannot be negative, these expressions reduce to the basic states that we have identified in Eqs. \eqref{b} and \eqref{bb}:
\begin{eqnarray}
|D^{1}_{0,0}M\rangle & =&  {b^\dagger_M} \nonumber \\
|D^{2}_{0,0}M\rangle &=& (\mathbf{b}^\dagger\mathbf{b}^\dagger)^2_M
\end{eqnarray}
The important conclusion from this overview of the D states is that all those states can be written as combinations of the basic vectors $|D^{1}_{0,0}M\rangle$ and $|D^{2}_{0,0}M\rangle$, with coefficients that only depend on the angular variable $x$ and the radial creation operator. This observation will allow us to construct the Ansatz.

As far as the F states are concerned, they are comparatively easy, since they only appear for $\nu=3p$, with $p\ge 1$. The fundamental F state thus results from the coupling of three bosons. Using the techniques explained earlier one can easily construct the component with the maximal $M=+3$ value. The other components can then easily be obtained by applying the $\mathcal{L}^-$ operator.
\begin{eqnarray}
(\mathbf{b}^\dagger\mathbf{b}^\dagger\mathbf{b}^\dagger)_{+3} &=& \frac{1}{2\sqrt{5}} \left[2 (b^\dagger_{+2})^2 b^\dagger_{-1} - \sqrt{6}\, b^\dagger_{+2}b^+_{+1}b^\dagger_{0} + (b^\dagger_{+1})^3 \right]
\nonumber \\
(\mathbf{b}^\dagger\mathbf{b}^\dagger\mathbf{b}^\dagger)_{+2} &=&\frac{1}{2\sqrt{30}} \left[4 (b^\dagger_{+2})^2 b^\dagger_{-2}+ 2\, b^\dagger_{+2} b^\dagger_{+1} b^\dagger_{-1} - 6 \, b^\dagger_{+2}(b^\dagger_0)^2 + \sqrt{6}\, (b^\dagger_{+1})^2 b^\dagger_{0} \right]\nonumber \\
(\mathbf{b}^\dagger\mathbf{b}^\dagger\mathbf{b}^\dagger)_{+1} &=& \frac{1}{2\sqrt{3}} \left[2 \, b^\dagger_{+2}b^\dagger_{+1} b^\dagger_{-2}+  ( b^\dagger_{+1})^2 b^\dagger_{-1} - \sqrt{6}\, b^\dagger_{+2} b^\dagger_{0} b^\dagger_{-1}\right] \nonumber \\
(\mathbf{b}^\dagger\mathbf{b}^\dagger\mathbf{b}^\dagger)_{0} &=& \frac{1}{2} \left[( b^\dagger_{+1})^2 b^\dagger_{-2}- b^\dagger_{+2}  ( b^\dagger_{-1})^2 \right] \nonumber \\
(\mathbf{b}^\dagger\mathbf{b}^\dagger\mathbf{b}^\dagger)_{-1} &=& \frac{1}{2\sqrt{3}} \left[-2 \, b^\dagger_{+2}b^\dagger_{-1} b^\dagger_{-2}-   b^\dagger_{+1} ( b^\dagger_{-1})^2 + \sqrt{6}\, b^\dagger_{+1} b^\dagger_{0} b^\dagger_{-2}\right] \nonumber \\
(\mathbf{b}^\dagger\mathbf{b}^\dagger\mathbf{b}^\dagger)_{-2} &=&\frac{1}{2\sqrt{30}} \left[-4\, b^\dagger_{+2} (b^\dagger_{-2})^2- 2\, b^\dagger_{+1} b^\dagger_{-1} b^\dagger_{-2} - 6 (b^\dagger_0)^2 b^\dagger_{-2} - \sqrt{6}\,b^\dagger_{0} (b^\dagger_{-1})^2  \right]\nonumber \\
(\mathbf{b}^\dagger\mathbf{b}^\dagger\mathbf{b}^\dagger)_{+3} &=& \frac{1}{2\sqrt{5}} \left[-2  b^\dagger_{-1}(b^\dagger_{-2})^2 + \sqrt{6}\,b^\dagger_{0} b^\dagger_{-1}b^+_{-2} - (b^\dagger_{-1})^3 \right]
\label{bbb}
\end{eqnarray}
Only two components of the octupole distortions have non-zero amplitude for the bimodal distortions:
\begin{equation}
\left(
\begin{matrix}
(\mathbf{b}^\dagger\mathbf{b}^\dagger\mathbf{b}^\dagger)_{+3} \\(\mathbf{b}^\dagger\mathbf{b}^\dagger\mathbf{b}^\dagger)_{+2} \\(\mathbf{b}^\dagger\mathbf{b}^\dagger\mathbf{b}^\dagger)_{+1}\\ (\mathbf{b}^\dagger\mathbf{b}^\dagger\mathbf{b}^\dagger)_0 \\(\mathbf{b}^\dagger\mathbf{b}^\dagger\mathbf{b}^\dagger)_{-1}\\ (\mathbf{b}^\dagger\mathbf{b}^\dagger\mathbf{b}^\dagger)_{-2}\\ (\mathbf{b}^\dagger\mathbf{b}^\dagger\mathbf{b}^\dagger)_{-3}
\end{matrix} \right)
=(\beta^{\dagger})^3 \frac{1}{2\sqrt{15}} 
\left(
\begin{matrix}
0\\
 -\sin{3\gamma} \\ 0 \\ 0\\ 0\\ \sin{3\gamma} \\ 0 
\end{matrix}
\right)
\end{equation}
Hence only the $F_{\pm2}$ components are non-zero. If we now introduce the spherical degrees of freedom, using the rotation matrix for the octupoles, $\mathbf{F}$, we obtain the general expressions for the F-modes in SO(5). The series starts with $p=1$:
\begin{equation}
|F_{1,0}M\rangle = (\mathbf{b}^\dagger\mathbf{b}^\dagger\mathbf{b}^\dagger)^3_M
=(\beta^\dagger)^3\sin{3\gamma} \left[ -F_{M2} + F_{M-2}\right]
\end{equation}
Here we have omitted the common normalizing factor.
The general expressions for the octupole states are given by Gheorgie et al. \cite{Gheorghe}: 
\begin{equation}
|F_{p,n}M\rangle = (\beta^\dagger)^{3p+2n} C^{3/2}_{p-1} (x)(\mathbf{b}^\dagger\mathbf{b}^\dagger\mathbf{b}^\dagger)^3_M
\end{equation}
%
%
\section{The Ansatz: vibronic P levels} 
As the Hamiltonian conserves SO(3) symmetry, all its solutions will be characterized by a definite angular momentum. The Ansatz is the result of the coupling between a fermionic part, which transforms as P, and a boson part. As we have indicated before there are no P-type bosons. From the coupling rules of angular momenta it is thus clear that the lowest possible vibronic state cannot be S, which could only result from the coupling of boson and fermion parts with the same P symmetry. The first vibronic states to be considered thus must have at least P symmetry. Such states arise through the coupling of the fermionic P with a bosonic S or D.
The $P\times D$ coupling is given by:
\begin{equation}
\left( \mathbf{f}^\dagger \mathbf{b}^\dagger \right)^1_q =
\sqrt{3} (-1)^{q-1} \left( \begin{matrix} 1&2&1 \\m_1& m_2 & -q \end{matrix} \right)
f^\dagger_{m_1} b^\dagger_{m_2} 
\end{equation}
For $q=+1$ this yield:
\begin{equation}
\left( \mathbf{f}^\dagger \mathbf{b}^\dagger \right)^1_1 = 
\frac{1}{\sqrt{10}} \left(f^\dagger_1 b^\dagger_0 - \sqrt{3}f^\dagger_0 b^\dagger_{+1} + \sqrt{6}f^\dagger_{-1} b^\dagger_{+2} \right)
\label{50}
\end{equation}
An entirely similar coupling arises for the two-boson modes, which also transform as D.
The Ansatz will thus consist of three parts, covering the S and the two D boson series.
Each of these parts involves an unknown function, $\mathcal{F}$, depending only on the radius of SO(5) space, and the angular SO(5) coordinate $x=\cos{3\gamma}$. The normalizer in Eq. \eqref{50} has been incorporated in these functions.
The Ansatz for the $M=+1, 0, -1$ components then reads:
\begin{eqnarray}
\Psi_{+1} &=&
\left(
\begin{matrix}
( \mathcal{F}_0 + b^\dagger_0 \mathcal{F}_1 + (\mathbf{b}^\dagger \mathbf{b}^\dagger )^2_0   \mathcal{F}_2 ) f^\dagger_{+1}\\ 
( -\sqrt{3} b^\dagger_{+1} \mathcal{F}_1 -\sqrt{3} (\mathbf{b}^\dagger \mathbf{b}^\dagger )^2_{+1}   \mathcal{F}_2 ) f^\dagger_{0}\\
( \sqrt{6} b^\dagger_{+2} \mathcal{F}_1 +\sqrt{6} (\mathbf{b}^\dagger \mathbf{b}^\dagger )^2_{+2}   \mathcal{F}_2 ) f^\dagger_{-1}
\end{matrix}
\right) \nonumber \\
\Psi_{0} &=&
\left(
\begin{matrix}
( \sqrt{3} b^\dagger_{-1} \mathcal{F}_1 + (\mathbf{b}^\dagger \mathbf{b}^\dagger )^2_{-1}   \mathcal{F}_2 ) f^\dagger_{+1}\\ 
(\mathcal{F}_0 -2 b^\dagger_{0} \mathcal{F}_1 -2 (\mathbf{b}^\dagger \mathbf{b}^\dagger )^2_{0}   \mathcal{F}_2 ) f^\dagger_{0}\\
(  \sqrt{3} b^\dagger_{+1} \mathcal{F}_1 +\sqrt{6} (\mathbf{b}^\dagger \mathbf{b}^\dagger )^2_{+1}   \mathcal{F}_2 ) f^\dagger_{-1}
\end{matrix}
\right) \nonumber \\
\Psi_{-1} &=&
\left(
\begin{matrix}
(  \sqrt{6} b^\dagger_{-2} \mathcal{F}_1 +\sqrt{6}  (\mathbf{b}^\dagger \mathbf{b}^\dagger )^2_{-2}   \mathcal{F}_2 ) f^\dagger_{+1}\\ 
( -\sqrt{3} b^\dagger_{-1} \mathcal{F}_1 -\sqrt{3} (\mathbf{b}^\dagger \mathbf{b}^\dagger )^2_{-1}   \mathcal{F}_2 ) f^\dagger_{0}\\
(\mathcal{F}_0 + b^\dagger_{0} \mathcal{F}_1 + (\mathbf{b}^\dagger \mathbf{b}^\dagger )^2_{0}   \mathcal{F}_2 ) f^\dagger_{-1}
\end{matrix}
\right) 
\end{eqnarray}
The spherical symmetry of the Hamiltonian is covered by the three angular degrees of freedom, which are embodied in the $LM_L$ angular momentum quantum numbers of the vibronic wavefunction. In other words, the Ansatz has SO(3) symmetry. Hence degeneracy of the three components of the  $\Psi$-wavefunction is procured.

To obtain the Jahn-Teller equations, we now have to apply the Bargmann mapping.
The creation operators are to be mapped on complex variables $z_{M}$. The conjugate annihilation operators have also conjugate symmetry properties, and are mapped on the derivatives. The radial creation operator is mapped on the complex variable $r$:
\begin{equation}
\beta^\dagger \rightarrow r
\end{equation}

Following the expressions in Eq. \eqref{tilde}, the Bargmann mapping of the coordinates is given by:
\begin{eqnarray}
Q_0  &\rightarrow& \frac{1}{\sqrt{2}} \left( z_0  + \frac{d}{dz_0}\right)
\nonumber \\
Q_{+1}  &\rightarrow& \frac{1}{\sqrt{2}} \left(  z_{+1}   -\frac{d}{dz_{-1}}\right)
\nonumber \\
Q_{-1}  &\rightarrow& \frac{1}{\sqrt{2}}  \left( z_{-1}   -\frac{d}{dz_{+1}}\right)
\nonumber \\
Q_{+2}  &\rightarrow& \frac{1}{\sqrt{2}} \left(  z_{+2}   + \frac{d}{dz_{-2}}\right)
\nonumber \\
Q_{-2}  &\rightarrow&  \frac{1}{\sqrt{2}} \left( z_{-2}  + \frac{d}{dz_{+2}}\right)
\end{eqnarray}

The essential variables of the boson space are the radius and the hyperspherical angle, $\gamma$. They are expressed as follows:
\begin{eqnarray}
r &=& \sqrt {z^2_0 - 2z_{+1}z_{-1} + 2 z_{+2}z_{-2}} \nonumber \\
x &=& \cos{3\gamma} \nonumber \\
&=& \frac{1}{r^3} \left[ z_0 (z_0^2 - 6 z_{+2}z_{-2} - 3 z_{+1}z_{-1})
+\frac{9}{\sqrt{6}} (z_{+2}z_{-1}^2 + z_{-2}z_{+1}^2) \right]
\end{eqnarray}
The 2-boson D-states in the ansatz are likewise expressed as follows:
\begin{eqnarray}
\left(\mathbf{b}^\dagger  \mathbf{b}^\dagger\right)_{0}^2 &\rightarrow& 
(\mathbf{z}\mathbf{z})^2_{0}=(z_0)^2  -2 z_{+2} z_{-2} - z_{+1}z_{-1}\nonumber \\
\left(\mathbf{b}^\dagger  \mathbf{b}^\dagger\right)_{\pm 1}^2 &\rightarrow& 
(\mathbf{z}\mathbf{z})^2_{\pm1} = z_0 z_{\pm 1} - \sqrt{6} z_{\pm 2}z_{\mp 1} \nonumber \\
\left(\mathbf{b}^\dagger \mathbf{b}^\dagger \right)_{\pm 2}^2 &\rightarrow& (\mathbf{z}\mathbf{z})^2_{\pm2} =\frac{\sqrt{3}}{\sqrt{2}} z_{\pm 1}^2 - 2 z_{\pm 2}z_{0} 
\end{eqnarray}
For all these quantities we also need the derivatives with respect to the $z$-variables. These are listed in Appendix A.

The harmonic part of the Hamiltonian has hyperspherical symmetry, and thus reduces to a purely radial number operator which simply counts the number of bosons:
\begin{equation}
\sum_m z_m\frac{d}{dz_m} = r\frac{d}{dr}
\end{equation}
The groundwork has now been laid, and we can finally apply the Hamiltonian to the Ansatz.
The crucial property on which the whole treatment rests, is that the $\mathcal{F}$ functions only depend on the variables $x$ and $r$.
\section{The Jahn-Teller equations}
 
When working out the Jahn-Teller equations various sums appear in the $z$-variables, which can conveniently be simplified using sum rules, which each time reduce complicated expressions to functions of $x,r$ and first powers of $z_m$ or $(\mathbf{z}\mathbf{z})_m^2$. These rules have been listed in Appendix B.

As a matrix operator, application of $\mathcal{H}$ on $|\Psi_{+1}\rangle$ yields three Jahn-Teller equations.
The first of these is obtained by combining the first row of the Hamiltonian with the column vector of the Ansatz:
\begin{equation}
\begin{split}
\left[ r\frac{d}{dr} - E+ \frac{\kappa}{\sqrt{6}}( z_0  + \frac{d}{dz_0} ) \right] & ( \mathcal{F}_0 + z_0 \mathcal{F}_1 + (\mathbf{z} \mathbf{z} )^2_0   \mathcal{F}_2 )
 \\
+  \frac{\kappa}{\sqrt{2}} \left( z_{-1}   -\frac{d}{dz_{+1}}\right) & ( -\sqrt{3} z_{+1} \mathcal{F}_1 -\sqrt{3} (\mathbf{z} \mathbf{z} )^2_{+1}   \mathcal{F}_2 ) \\
+\kappa \left( z_{-2}  + \frac{d}{dz_{+2}}\right) & ( \sqrt{6} z_{+2} \mathcal{F}_1 +\sqrt{6} (\mathbf{z} \mathbf{z} )^2_{+2}   \mathcal{F}_2 ) = 0 \\
\end{split}
\end{equation}
This equation can be rewritten as:
\begin{equation}
\mathcal{G}_0 + z_0 \mathcal{G}_{1} +(\mathbf{z} \mathbf{z} )^2_0 \mathcal{G}_2 =0
\label{57}
\end{equation}
where:
\begin{equation}
\begin{split}
\mathcal{G}_0 =& (r\frac{d}{dr} - E) \mathcal{F}_0 + \frac{2\kappa}{\sqrt{6}} \left[5 +  r(r + \frac{\delta }{\delta r}) \right]\mathcal{F}_1 \\
&+
\kappa \left[\frac{2}{\sqrt{6}} xr^2(r + \frac{\delta }{\delta r}) + \sqrt{6} r(1-x^2)\frac{\delta}{\delta x}\right] \mathcal{F}_2 \\
\end{split}
\end{equation}

\begin{equation}
\begin{split}
\mathcal{G}_1 = &\frac{\kappa}{\sqrt{6}} \left(1 +\frac{1}{r} \frac{\delta}{\delta r} - 3\frac{x}{r^2}\frac{\delta}{\delta x}\right) \mathcal{F}_0 \\
&+\left[r\frac{\delta}{\delta r} - E +1 -\frac{3\kappa}{\sqrt{6}}\frac{1}{r} \frac{\delta}{\delta x}\right]\mathcal{F}_1 \\
& -\frac{\kappa}{\sqrt{6}}\left[ 7 + r(r + \frac{\delta}{\delta r}) + 3x\frac{\delta}{\delta x}\right] \mathcal{F}_2 \\
\end{split}
\end{equation}

\begin{equation}
\begin{split}
\mathcal{G}_2 = & \frac{3\kappa}{\sqrt{6}} \frac{1}{r^3} \frac{\delta}{\delta x} \mathcal{F}_0 
-\frac{\kappa}{\sqrt{6}}\left(1 +\frac{1}{r} \frac{\delta}{\delta r} - 3\frac{x}{r^2}\frac{\delta}{\delta x}\right)\mathcal{F}_1 \\
&+ \left( r\frac{\delta}{\delta r} - E + 2 + \frac{3\kappa}{\sqrt{6}}\frac{1}{r} \frac{\delta}{\delta x}\right) \mathcal{F}_2 \\
\end{split}
\end{equation}
Applying now the second and third row of the Hamiltonian to the $|\Psi_1\rangle$ column vector yields:
\begin{eqnarray}
-\sqrt{3} [ z_{+1} \mathcal{G}_{1} + (\mathbf{z} \mathbf{z} )^2_{+1} \mathcal{G}_2] & =& 0
\nonumber \\
\sqrt{6} [ z_{+2} \mathcal{G}_{1} + (\mathbf{z} \mathbf{z} )^2_{+2} \mathcal{G}_2] & =& 0
\label{61}
\end{eqnarray}
This is a gratifying result. Since the $z$ variables are linearly independent, Eqs. \eqref{57} and \eqref{61} can only be fulfilled if all $\mathcal{G}$ functionals be equal to zero. Thus one obtains finally three Jahn-Teller equations:
\begin{eqnarray}
\mathcal{G}_0 &=& 0\nonumber \\
\mathcal{G}_1 &=&0 \nonumber \\
\mathcal{G}_2 &=& 0
\end{eqnarray}
The same conclusion is verified when applying the Hamiltonian to $|\Psi_0\rangle$ or $|\Psi_{-1}\rangle$, thus demonstrating once again the SO(3) invariance of the triplet Jahn-Teller problem.
\section{Solution by series expansion}
To work out the $\mathcal{G}_i$ equations, we expand the $\mathcal{F}$ functionals in the SO(5) harmonics, using three sets of coefficients: $a$ for the S states, $b$ and $c$ for the D states. 
\begin{eqnarray}
\mathcal{F}_0 &=&  \sum_{n,p} a^n_p r^{2n+3p} C_p^{1/2}(x)
\nonumber \\
\mathcal{F}_1 &=& \sum_{n,p} b^n_p r^{2n+3p} C_p^{3/2} - \sum_{n,p} c^n_{p} r^{1+2n+3p} C_{p-1}^{3/2} \nonumber \\
\mathcal{F}_2 &=& -\sum_{n,p} b^n_p r^{2n+3p-1} C_{p-1}^{3/2} + \sum_{n,p} c^n_p r^{2n+3p} C_{p}^{3/2}
\end{eqnarray}
The summations in this equations are limited by the validity range of the Gegenbauer polynomials, the rank of which can not be lower than zero. Hence if a polynomial appears of rank $p-1$ the lowest allowed value of $p$ is $+1$.
The $\mathcal{F}_0$ function is an expansion in the S-states.
\begin{equation}
\mathcal{F}_0 =  \sum_{n,p} a^n_p |S_{p,n}\rangle
\end{equation}
The case of the $\mathcal{F}_1$ and $\mathcal{F}_2$ functions is peculiar, since they are based on the same sets of coefficients. Note that in the SU(5) spectrum, there are two sets of D states, depending on whether the seniority label is $3\nu+1$ or $3\nu + 2$. The wavefunction should contain both sets and run independent expansions in both sets. This is why we need $b$ and $c$ type expansion coefficients. Indeed when we combine the $\mathcal{F}_1$ and $\mathcal{F}_2$ functions as defined by the Ansatz, we obtain precisely the required twofold series expansion for the D states:
\begin{equation}
b^{\dagger}_M \mathcal{F}_1 + {(\mathbf{b}^\dagger\mathbf{b}^\dagger)_M}\mathcal{F}_2
= \sum_{n,p} b^n_p |D^{1}_{p,n}M \rangle + \sum_{n,p} c^n_p |D^{2}_{p,n}M \rangle
\end{equation}
The operators in the equations are now applied to these functions. This is a very cumbersome but straightforward procedure. In this process several identities of Gegenbauer polynomials are useful \cite{Szego}. The derivative of a Gegenbauer raises the dimension but lowers the rank. In contrast multiplication by $x$ will partly lower the dimension:
\begin{eqnarray}
\frac{d}{dx} C^{\lambda}_p(x) &=&2\lambda C^{\lambda+1}_{p-1}(x) \nonumber \\
x C^{\lambda}_p(x) &=& \frac{p+1}{2(\lambda-1)}C^{\lambda-1}_{p+1}(x)+C^{\lambda}_{p-1}(x) \nonumber \\
x\frac{d}{dx} C^{\lambda}_p(x) &=&p C^{\lambda}_{p}(x)+2\lambda C^{\lambda+1}_{p-2}(x)
\end{eqnarray}
Further, the Jahn-Teller equations also contain one operator which conserves the dimension:
\begin{equation}
(1-x)^2 \frac{d}{dx}C^{\lambda}_p(x) = \frac{1}{2(p+\lambda)}
\left[(p+2\lambda-1)(p+2\lambda)C^{\lambda}_{p-1}(x)-p(p+1) C^{\lambda}_{p+1}(x)\right]
\end{equation}
The raising or lowering of the dimensions could be cause of serious concern, as orthogonality of Gegenbauer polynomials only applies to polynomials of the same dimension. So in the end all Gegenbauer polynomials should be of equal dimension. In fact the Jahn-Teller equations produce Gegenbauer polynomials of dimensions $1/2$, $3/2$ and $5/2$. In order to be able to solve the equations, it is mandatory that all these can be reduced to polynomials of dimension $3/2$. As it happens, there are some fortunate coincidences which remove initial concerns. To polynomials with $\lambda=1/2$ the following identity can be applied to \emph{raise} the dimension to $\lambda=3/2$:
\begin{equation}
C_p^{\lambda} = \frac{\lambda}{p+\lambda} \left( C_p^{\lambda + 1} - C_{p-2}^{\lambda + 1}\right)
\end{equation}
When applying this identity, keep in mind that polynomials $C_p^{\lambda}$ with rank $p<0$ are zero.
A relationship of this type can however not simply be inverted to \emph{reduce} the dimension by one unit, as it will work only for the exact difference of polynomials with ranks $p$ and $p-2$.
In the case of $\mathcal{G}_1$ and $\mathcal{G}_2$, by a touch of magic the Gegenbauer polynomials of dimension $5/2$ can be eliminated, due to two fortunate combinations of operators.
In one case the terms are subtracted precisely in the way required to reduce their dimension to the $3/2$.
\begin{equation}
\begin{split}
-\frac{1}{r}\frac{\delta}{\delta x} & r^{2n+3p}C^{3/2}_{p} + x\frac{\delta}{\delta x} r^{2n+3p-1}C^{3/2}_{p-1}\\
&= r^{2n+3p-1} \left[ -\frac{\delta}{\delta x} C^{3/2}_p+x\frac{\delta}{\delta x} C^{3/2}_{p-1}\right] \\
&=  r^{2n+3p-1} \left[ (p-1) C^{3/2}_{p-1} - 3 (C^{5/2}_{p-1}- C^{5/2}_{p-3})\right] \\
&= -r^{2n+3p-1}(p+2)C_{p-1}^{3/2}
\end{split}
\end{equation}
In the other case, the terms of dimension $5/2$ simply cancel:
\begin{equation}
\begin{split}
-\frac{1}{r}\frac{\delta}{\delta x} & r^{2n+3p+1}C^{3/2}_{p-1} + x\frac{\delta}{\delta x} r^{2n+3p}C^{3/2}_{p}\\
&= r^{2n+3p} \left[ -\frac{\delta}{\delta x} C^{3/2}_{p-1}+x\frac{\delta}{\delta x} C^{3/2}_{p}\right] \\
&=  r^{2n+3p} \left[ - 3 C^{5/2}_{p-2}+ p C^{3/2}_p +3 C^{5/2}_{p-2}\right] \\
&= r^{2n+3p}p C_{p}^{3/2}
\end{split}
\end{equation}
We are thus able to reduce all equations to recurrence formulas based on powers of $r$, and Gegenbauer polynomials of dimension $3/2$. 
\begin{equation}
\begin{split}
\mathcal{G}_0 &= \sum_{n,p} a^n_p \frac{2n+3p-E}{2p+1}  r^{2n+3p}( C_p^{3/2}-C_{p-2}^{3/2}) \\
&+\frac{2\kappa}{\sqrt{6}}\sum_{n,p} b^n_p  \left[ (2n + 3p+5)  r^{2n+3p} +  r^{2n+3p+2} \right]C_p^{3/2} \\
& - \frac{2\kappa}{\sqrt{6}} \sum_{n,p} b^n_p \frac{1}{2p+1}\left[(2n+3p-1)r^{2n+3p} +  r^{2n+3p+2}\right] \left[pC_p^{3/2} +(p+1)C_{p-2}^{3/2} \right]  \\
& -\kappa\sqrt{6} \sum_{n,p} b^n_p r^{2n+3p} \frac{1}{2p+1} \left[ (p+1)(p+2)C_{p-2}^{3/2} - p(p-1) C_{p}^{3/2} \right]  \\
&-\frac{2\kappa}{\sqrt{6}}\sum_{n,p} c^n_p  \left[ (
2n + 3p+6)  r^{2n+3p+1} +  r^{2n+3p+3} \right]C_{p-1}^{3/2} \\
& +\frac{2\kappa}{\sqrt{6}} \sum_{n,p} c^n_p \frac{2n+3p}{2p+3}(r^{2n+3p+1} +  r^{2n+3p+3}) \left [(p+1)C_{p+1}^{3/2} +(p+2)C_{p-1}^{3/2} )\right] \\
& +\kappa\sqrt{6} \sum_{n,p} c^n_p r^{2n+3p+1} \frac{1}{2p+3} \left[ (p+2)(p+3)C_{p-1}^{3/2} - p(p+1) C_{p+1}^{3/2} \right]  \\
\\
\mathcal{G}_1 & = \frac{\kappa}{\sqrt{6}} \sum_{n.p} a^n_p \frac{1}{2p+1} r^{2n+3p} \left[C_p^{3/2}-C_{p-2}^{3/2}\right] \\
&+\frac{\kappa}{\sqrt{6}} \sum_{n.p} a^n_p \frac{1}{2p+1}  r^{2n+3p-2} \left[2n C_p^{3/2}-(2n+6p+3)C^{3/2}_{p-2}\right] \\
&+ \sum_{n,p} b^n_p (2n+3p +1-E) r^{2n+3p}C_p^{3/2} \\
&+ \frac{\kappa}{\sqrt{6}} \sum_{n,p} b^n_p ( 2n r^{2n+3p-1} + r^{2n+3p+1} ) C_{p-1}^{3/2} \\
& -\sum_{n,p} c^n_p (2n+3p +2-E) r^{2n+3p+1}C_{p-1}^{3/2} \\
&- \frac{\kappa}{\sqrt{6}} \sum_{n,p} c^n_p \left[ (2n +6p +7)r^{2n+3p} + r^{2n+3p+2} \right] C_{p}^{3/2}
\end{split}
\end{equation}
\newpage
\begin{equation}
\begin{split}
\mathcal{G}_2 &=\frac{3\kappa}{\sqrt{6}} \sum_{n,p} a^n_p r^{2n+3p-3} C_{p-1}^{3/2} \\
&-\frac{\kappa}{\sqrt{6}}\sum_{n,p}b^n_p ( r^{2n+3p} + 2nr^{2n+3p-2}) C_p^{3/2} \\
&- \sum_{n,p} b^n_p (2n+3p+1 -E) r^{2n+3p-1} C_{p-1}^{3/2} \\
&+\frac{\kappa}{\sqrt{6}}\sum_{n,p} c^n_p ( r^{2n+3p+1} + (2n+6p+7)r^{2n+3p-1}) C_{p-1}^{3/2} \\
&+ \sum_{n,p} c^n_p (2n+3p+2 -E) r^{2n+3p} C_{p}^{3/2}
\end{split} 
\end{equation}

\section{Example}
From each of the Jahn-Teller equations one can derive recurrence relationships by isolating all terms with the same power of $r$ ánd the same rank of the Gegenbauer polynomial. These relations may then be collected in a matrix equation, which is truncated at a certain highly excited oscillator level. 
As an example we consider the solution within a space for the six lowest S and D states in Figure 1. The basis can be arranged in a vector as; $a_0^0,b^0_0, a^1_0, c_0^0, b^1_0,a_1^0 $. Collecting terms in the lowest powers of r and the lowest Gegenbauer ranks, yields six equations in these variables, which are ordered in matrix form as:
\begin{equation}
\left(
\begin{matrix}
- E & \frac{10\kappa}{\sqrt{6}}& 0 & 0 &0 &0 \\
\frac{\kappa}{\sqrt{6}} & 1-E & \frac{2\kappa}{\sqrt{6}}k & -\frac{7\kappa}{\sqrt{6}} & 0 & 0 \\
0& \frac{2\kappa}{\sqrt{6}} & 2-E & 0 & \frac{14\kappa}{\sqrt{6}} & 0 \\
0 & -\frac{\kappa}{\sqrt{6}}k & 0 & 2-E &-\frac{2\kappa}{\sqrt{6}}k & \frac{3\kappa}{\sqrt{6}} \\
0 & 0 & \frac{\kappa}{\sqrt{6}} & -\frac{\kappa}{\sqrt{6}} & 3-E & 0 \\
0 & 0 & 0& \frac{2\kappa}{\sqrt{6}} & 0 &3-E 
\end{matrix}
\right)
\left(
\begin{matrix}
a_0^0 \\
b^0_0 \\
a^1_0 \\
c_0^0 \\
b^1_0 \\
a^0_1
\end{matrix}
\right)
=0
\end{equation}
For $\kappa=1$ the secular equation of degree 6 is given by:
\begin{equation}
18 x^6 -198x^5 +717 x^4 -786 x^3 -612 x^2 +1439 x -490=0
\end{equation}
This equation corresponds exactly to the sextic equation that results from the group-theoretical treatment by Judd, with force element $k=\sqrt{5}\kappa$ \cite{Judd3}. The roots, in units of $\hbar\omega$, are: -1.1597, +0.4697, +1.3736, +2.2853, +3.5630, +4.4680. 
\section{Conclusions}
The Bargmann mapping offers a compact analytical method to describe oscillator problems. Reik has initiated the application of this method in the area of boson-fermion interactions. So far only the simplest vibronic coupling problems were described, viz. the Rabi Hamiltonian and the generalized $E\times e$ Jahn-Teller Hamiltonian \cite{Reik}. In each case the essential part of the solution is to construct an Ansatz. An Ansatz is a template which provides the general structure of the wavefunction. It must comply with the symmetry requirements and lead to an equation which is entirely based on the invariants of the interaction. 
The method can also be applied to the $\Gamma_8$ quartet instability, but, apparently, so far there are no reports of application to the $T$ triplet instability. The principal aim of our study has thus been to obtain the missing Ansatz. Since in this case there is no common symmetry group which covers both the fermion and the boson parts, symmetry breaking had to be explicitly taken into account when constructing the Ansatz. Also closed formula were required for the hyperspherical quadrupolar oscillator states. These were provided by the finite expansions of these states based on Gegenbauer polynomials by Gheorghe et al. \cite{Gheorghe}.

In principle the present method can be extended to any $L$ value. The Jahn-Teller equations will always provide sufficient recursion formulas, due to the presence of separate orthogonality requirements for the radial and angular basis functions. In this way it should be possible to tackle some intricate multiplicity problems, which appear for states with $L\ge 6$.
Further studies will explore the potential of this method to elucidate the quantization requirements for boson-fermion couplings, and to look for new Juddian exact solutions \cite{Judd4}.
\newpage
\appendix
\begin{appendices}
\section{A. Derivatives of the $z$-functions}

\begin{tabular}{l|cc}
 \vspace{3mm}
 & $r$ & $x$ \\
 \hline
 \vspace{3mm}
 $\frac{d}{dz_0}$ & $\frac{z_0}{r}$&$\frac{3(\mathbf{z}\mathbf{z})^2_{0}}{r^3} -\frac{3xz_0}{r^2}$  \\
 \vspace{3mm}
  $\frac{d}{dz_{\pm1}}$ &$-\frac{z_{\mp1}}{r}$ & $-\frac{3 (\mathbf{z}\mathbf{z})^2_{\mp1}}{r^3} + \frac{3xz_{\mp1}}{r^2}$\\
 \vspace{3mm}
$\frac{d}{dz_{\pm2}}$ & $\frac{z_{\mp2}}{r}$& $\frac{3 (\mathbf{z}\mathbf{z})^2_{\mp2}}{r^3} - \frac{3xz_{\mp2}}{r^2}$
\end{tabular}
%
\vspace{1cm}

%
\begin{tabular}{l|cccccc}
 \vspace{3mm}
 & $(\mathbf{z}\mathbf{z})^2_{0}$ & $(\mathbf{z}\mathbf{z})^2_{+1}$ &$(\mathbf{z}\mathbf{z})^2_{-1}$ &$(\mathbf{z}\mathbf{z})^2_{+2}$ &$(\mathbf{z}\mathbf{z})^2_{-2}$\\
 \hline
 \vspace{3mm}
 $\frac{d}{dz_0}$ & $2{z_0}$ & $-z_{+1}$ & $-z_{-1}$ & $-2z_{+2}$ & $-2z_{-2}$ \\
 \vspace{3mm}
 $\frac{d}{dz_{+1}}$ &$-z_{-1}$ & $z_0$&$-\sqrt{6}z_{-2}$& $\sqrt{6} z_{+1}$&$0$\\
    \vspace{3mm}
$\frac{d}{dz_{-1}}$ &$-z_{+1}$  &$-\sqrt{6}z_{+2}$&$z_0$& $0$ & $\sqrt{6} z_{-1}$\\
 \vspace{3mm}
$\frac{d}{dz_{+2}}$ & $-2z_{-2}$& $-\sqrt{6} z_{-1}$ & $0$ & $-2z_0$ & $0$\\
 \vspace{3mm}
$\frac{d}{dz_{-2}}$ & $-2z_{+2}$& $0$ & $-\sqrt{6}z_{+1}$ & $0$ & $-2z_0$
\end{tabular}
\section{B. Sum rules involving polynomials of the $z$variables}
\begin{equation}
(z_0)^2 -3z_{+1}z_{-1}+6z_{+2}z_{-2} = 2r^2 - (\mathbf{z}\mathbf{z})_0^2
\end{equation}
\begin{eqnarray}
z_0(\mathbf{z}\mathbf{z})_0^2-3z_{\pm1}(\mathbf{z}\mathbf{z})_{\mp1}^2+6z_{\pm2}(\mathbf{z}\mathbf{z})_{\mp2}^2 &=&2xr^3 -z_0r^2  \nonumber \\
- z_0(\mathbf{z}\mathbf{z})_{+1}^2 + 2z_{+1}(\mathbf{z}\mathbf{z})_0^2 -\sqrt{6} z_{+2} (\mathbf{z}\mathbf{z})_{-1}^2 &=& z_{+1}r^2 \nonumber \\
- z_{+1}(\mathbf{z}\mathbf{z})_{0}^2 + 2z_{0}(\mathbf{z}\mathbf{z})_{+1}^2 -\sqrt{6} z_{-1} (\mathbf{z}\mathbf{z})_{+2}^2 &=& z_{+1}r^2 \nonumber \\
 z_0(\mathbf{z}\mathbf{z})_{+2}^2 - \sqrt{3/2}z_{+1}(\mathbf{z}\mathbf{z})_{+1}^2 + z_{+2} (\mathbf{z}\mathbf{z})_{0}^2 &=&- z_{+2}r^2 
 \end{eqnarray}
 \begin{eqnarray}
 \left[(\mathbf{z}\mathbf{z})_{0}^2\right]^2 -3(\mathbf{z}\mathbf{z})_{+1}^2(\mathbf{z}\mathbf{z})_{-1}^2+6 (\mathbf{z}\mathbf{z})_{+2}^2(\mathbf{z}\mathbf{z})_{-2}^2
 &=&-2z_0xr^3 +(\mathbf{z}\mathbf{z})_{0}^2 r^2 + 2r^4 \nonumber \\
 2(\mathbf{z}\mathbf{z})_{0}^2(\mathbf{z}\mathbf{z})_{+2}^2- \sqrt{3/2} \left[(\mathbf{z}\mathbf{z})_{+1}^2\right]^2 &=& -2z_{+2} xr^3 + (\mathbf{z}\mathbf{z})_{+2}^2r^2 \nonumber \\
 (\mathbf{z}\mathbf{z})_{0}^2(\mathbf{z}\mathbf{z})_{+1}^2
-\sqrt{6} (\mathbf{z}\mathbf{z})_{+2}^2(\mathbf{z}\mathbf{z})_{-1}^2 
 &=& 2z_{+1}xr^3 - (\mathbf{z}\mathbf{z})_{+1}^2r^2
\end{eqnarray}
\end{appendices}
\newpage
\bibliography{AnsatzTx(e+t)}

\begin{thebibliography}{24}
\providecommand{\natexlab}[1]{#1}
\providecommand{\url}[1]{\texttt{#1}}
\expandafter\ifx\csname urlstyle\endcsname\relax
  \providecommand{\doi}[1]{doi: #1}\else
  \providecommand{\doi}{doi: \begingroup \urlstyle{rm}\Url}\fi

\bibitem[Arvanitidis et~al.(2017)Arvanitidis, Vandaele, Szopa, and
  Ceulemans]{Ceulemans2}
A.~Arvanitidis, E.~R.~J. Vandaele, M.~Szopa, and A.~Ceulemans.
\newblock {The quantization of the $E\times e$ Jahn-Teller Hamiltonian}.
\newblock \emph{J. Phys. Chem. A}, 121:\penalty0 7246--7254, 2017.

\bibitem[Bargmann(1961)]{Bargmann}
V.~Bargmann.
\newblock {On a Hilbert space of analytic functions and an associated integral
  transform}.
\newblock \emph{Communications on Pure and Applied Mathematics}, XIV:\penalty0
  187--214, 1961.

\bibitem[Bersuker(2006)]{Bersuker}
I.~B. Bersuker.
\newblock \emph{The Jahn-Teller effect}.
\newblock Cambridge University Press, Cambridge, 2006.

\bibitem[Bes(1959)]{Bes}
D.~R. Bes.
\newblock The $\gamma$-dependent part of the wave functions representing
  $\gamma$-unstable surface vibrations.
\newblock \emph{Nuclear Physics}, 10:\penalty0 373--385, 1959.

\bibitem[Braak(2011)]{Braak}
D.~Braak.
\newblock {Integrability of the Rabi model}.
\newblock \emph{Phys. Rev. Lett.}, 107:\penalty0 100401, 2011.

\bibitem[Ceulemans et~al.(1994)Ceulemans, Fowler, and Vos]{Ceulemans3}
A.~Ceulemans, P.~W. Fowler, and I.~Vos.
\newblock {$C_{60}$ vibrates as a hollow sphere}.
\newblock \emph{J. Chem. Phys.}, 100:\penalty0 5491--5500, 1994.

\bibitem[Chac\'{o}n and Moshinsky(1977)]{Chacon}
E.~Chac\'{o}n and M.~Moshinsky.
\newblock Group theory of the collective model of the nucleus.
\newblock \emph{J. Math. Phys.}, 18:\penalty0 870--880, 1977.

\bibitem[Chancey and Judd(1983)]{Judd3}
C.~C. Chancey and B.~R. Judd.
\newblock {An approximate analytical treatment of the $T\times(e+t_2)$
  Jahn-Teller effect}.
\newblock \emph{J. Phys. A: Math. Gen.}, 16:\penalty0 875--890, 1983.

\bibitem[Chancey and O'Brien(1997)]{Chancey}
C.~C. Chancey and M.~C.~M. O'Brien.
\newblock \emph{{The Jahn-Teller effect in $C_{60}$ and other icosahedral
  complexes}}.
\newblock Princeton University Press, Princeton, New Jersey, 1997.

\bibitem[Eisenberg and Greiner(1987)]{Eisenberg}
J.~M. Eisenberg and W.~Greiner.
\newblock \emph{Nuclear models: collective and single particle phenomena}.
\newblock North Holland, Amsterdam, 1987.

\bibitem[Englman and Yahalom(2001)]{Englman}
R.~Englman and A.~Yahalom.
\newblock {The Jahn-Teller effect: a permanent presence in the frontiers of
  science}.
\newblock In M.~D. Kaplan and G.~O. Zimmerman, editors, \emph{Vibronic
  interactions: Jahn-Teller effect in crystals and molecules. Nato Science
  Series}, volume II39, pages 5--14. Kluwer, Dordrecht, 2001.

\bibitem[Gheorghe et~al.(1978)Gheorghe, Raduta, and Ceausescu]{Gheorghe}
A.~Gheorghe, A.~A. Raduta, and V.~Ceausescu.
\newblock On the exact solution of the harmonic quadrupole collective
  hamiltonian.
\newblock \emph{Nuclear Physics}, A296:\penalty0 228--250, 1978.

\bibitem[Judd(1974)]{Judd2}
B.~R. Judd.
\newblock {Lie groups and the Jahn-Teller effect}.
\newblock \emph{Can. J. Phys.}, 52:\penalty0 999--1044, 1974.

\bibitem[Judd(1979)]{Judd4}
B.~R. Judd.
\newblock {Exact solutions to a class of Jahn-Teller systems}.
\newblock \emph{J. Phys. C: Solid State Phys.}, 12:\penalty0 1685--1692, 1979.

\bibitem[Judd(1987)]{Judd1}
B.~R. Judd.
\newblock Group theoretical approaches.
\newblock In Yu.~E. Perlin and M.~Wagner, editors, \emph{{The dynamical
  Jahn-Teller effect in localized systems}}, chapter~3. Elsevier, Amsterdam,
  1987.

\bibitem[LeTourneux(1965)]{Tourneux}
J.~LeTourneux.
\newblock Effect of the dipole-quadrupole interaction on the width and the
  structure of the giant dipole line in spherical nuclei.
\newblock \emph{Mat. Fys. Med. Kon. Danske Videns. Selskab.}, 34\penalty0 (11),
  1965.

\bibitem[Liu et~al.(2018)Liu, Iwahara, and Chibotaru]{Naoya}
D.~Liu, N.~Iwahara, and L.~F. Chibotaru.
\newblock {Dynamical Jahn-Teller effect of fullerene anions}.
\newblock \emph{Phys. Rev. B}, 97:\penalty0 115412, 2018.

\bibitem[O'Brien(1971)]{Brien}
M.~C.~M. O'Brien.
\newblock {The Jahn-Teller effect in a P state equally coupled to $E_g$ and
  ${T_2}_g$ vibrations}.
\newblock \emph{J. Phys. C: Solid State Phys.}, 4:\penalty0 2524--2536, 1971.

\bibitem[Reik(1987)]{Reik}
H.~G. Reik.
\newblock {The Jahn-Teller effect: a permanent presence in the frontiers of
  science}.
\newblock In Yu.~E. Perlin and M.~Wagner, editors, \emph{The dynamical
  Jahn-Teller effect in localized systems}, chapter~4. Elsevier, Amsterdam,
  1987.

\bibitem[Rowe et~al.(2004)Rowe, Turner, and Repka]{Rowe1}
D.~J. Rowe, P.~S. Turner, and J.~Repka.
\newblock {Spherical harmonics and basic coupling coefficients for the group
  SO(5) in an SO(3) basis}.
\newblock \emph{J. Math. Phys.}, 45:\penalty0 2761--2784, 2004.

\bibitem[Rowe and Wood(2010)]{Rowe2}
D.~R. Rowe and J.~L. Wood.
\newblock \emph{Fundamentals of nuclear models}.
\newblock World Scientific, Singapore, 2010.

\bibitem[Szegö(1975)]{Szego}
G.~Szegö.
\newblock \emph{{Orthogonal polynomials, 4th Ed}}.
\newblock American Mathematical Society, Providence, Rhode Island, 1975.

\bibitem[Thyssen and Ceulemans(2017)]{Ceulemans4}
P.~Thyssen and A.~Ceulemans.
\newblock \emph{Shattered symmetry: group theory from the eightfold way to the
  periodic table}.
\newblock Oxford University Press, Oxford, 2017.

\bibitem[Vandaele et~al.(2017)Vandaele, Arvanitidis, and Ceulemans]{Ceulemans1}
E.~R.~J. Vandaele, A.~Arvanitidis, and A.~Ceulemans.
\newblock {The quantization of the Rabi Hamiltonian}.
\newblock \emph{J. Phys. A: Math. Theor.}, 50:\penalty0 114002, 2017.

\end{thebibliography}

\end{document}